\documentclass[12pt,preprint]{aastex}

\slugcomment{Submitted to ApJ Letters}
\shorttitle{Turbulent Warm Ionized Medium}
\shortauthors{Wu et al.}

\font\mbf = cmmib10 scaled\magstep1
       \font\mbfs = cmmib10 \font\mbfss = cmmib10 scaled 833
       \def\bmit{\fam9 }

\begin{document}
\title{Estimation of Magnetic Field Strength in the Turbulent
Warm Ionized Medium}

\author{Qingwen Wu\altaffilmark{1}, Jongsoo Kim\altaffilmark{1,2},
Dongsu Ryu\altaffilmark{3,4}, Jungyeon Cho\altaffilmark{3}, and
Paul Alexander\altaffilmark{2}}

\altaffiltext{1}{Korea Astronomy and Space Science Institute,
61-1, Hwaam-Dong, Yuseong-Gu, Daejeon 305-348, South Korea;
qwu@kasi.re.kr}
\altaffiltext{2}{Astrophysics Group, Cavendish Laboratory,
Cambridge University, JJ Thompson Avenue, Cambridge CB3 0HE, UK;
jskim@mrao.cam.ac.uk, pa@mrao.cam.ac.uk}
\altaffiltext{3}{Department of Astronomy and Space Science,
Chungnam National University, Daejeon 305-764, South Korea;
ryu@canopus.cnu.ac.kr, cho@canopus.cnu.ac.kr}
\altaffiltext{4}{Corresponding Author}

\begin{abstract}

We studied Faraday rotation measure (RM) in turbulent media with the rms
Mach number of unity, using isothermal, magnetohydrodynamic turbulence
simulations.
Four cases with different values of initial plasma beta were considered.
Our main findings are as follows.
(1) There is no strong correlation between the fluctuations of magnetic
field strength and gas density.
So the magnetic field strength estimated with RM/DM (DM is the dispersion
measure) correctly represents the true mean strength of the magnetic
field along the line of sight.
(2) The frequency distribution of RMs is well fitted to the Gaussian.
In addition, there is a good relation between the width of the
distribution of RM/$\overline{\rm RM}$ ($\overline{\rm RM}$ is the
average value of RMs) and the strength of the regular field along the
line of sight; the width is narrower, if the field strength is stronger.
We discussed the implications of our findings in the warm ionized medium
where the Mach number of turbulent motions is around unity.

\end{abstract}


\section{Introduction}

The warm ionized medium (WIM) is one of the major gas components of
our Galaxy.
It is a diffuse ionized gas with temperature $T \sim 8000$ K, scale
height $H \sim 1$ kpc, and average density ${\bar n} \sim 0.03$ cm$^{-3}$,
and occupies approximately 20\% of the volume of the disk in the Galaxy
\citep[e.g.,][]{reyn91,ha99}.
The measured width of the H$\alpha$ line from the WIM is in the range of
15 to 50 km s$^{-1}$ \citep{trh99}.
Observations indicate that the WIM is in a turbulent state (see below).
Assuming that the typical width of H$\alpha$ line and temperature are
30 km s$^{-1}$ and 8000 K, respectively, the non-thermal, turbulent
velocity would be about 13 km s$^{-1}$, as is calculated, for instance,
with Equation (1) of \citet{reyn85}.
So the sonic Mach number of turbulent motions, $M_s$, would be around
unity in the WIM.
It is smaller than $M_s$ of the cold neutral medium, which is a few
\citep[e.g.,][]{ht03}, and $M_s$ of molecular clouds, which is
$\ga 10$ \citep[e.g.,][]{la81}.

The best evidence for turbulence in the interstellar medium (ISM) which
includes the WIM comes from the power spectrum presented in \citet{ar95}.
It is a composite power spectrum of electron density collected from
observations of velocity fluctuations of the interstellar gas, rotation
measures (RM), dispersion measures (DM), interstellar scintillations,
and others.
The spectrum covers the spatial range of $\sim 10^{10}-10^{20}$ cm,
and remarkably the whole spectrum is fitted to the power spectrum of
Kolmogorov turbulence with slope $-5/3$.
In addition, it has been recently reported that the H$\alpha$ emission
measure for the WIM \citep{hi08} and the densities of the diffuse
ionized gas and the diffuse atomic gas \citep{bf08} follow the lognormal
distribution.
The lognormality in density (or column density) distributions can be
regarded as another signature of turbulence in the WIM
\citep[e.g.,][and references therein]{vs94,es04}.

The interstellar magnetic field that is pervasive in the Galaxy
plays an important role in the dynamics of the ISM, star formation,
acceleration of cosmic rays, etc.
It has the energy density comparable to those of turbulence and
cosmic rays, as well as the thermal energy density \citep{sp78}.
The information on the field has been obtained through observations
of Zeeman splitting, polarized thermal emission from dust, optical
starlight polarization, radio synchrotron emission, and the Faraday
rotation of polarized radio sources \citep[see,][for a review]{hw02}.
Of them, the last method can be used to study the magnetic field in
ionized media such as the WIM.
It estimates the mean strength of the magnetic field along the line
of sight, weighted with the electron density $n_e$,
\begin{equation}
\left<B_{\parallel}\right> = \frac{\int n_{e} B_{\parallel} ds}{\int n_e ds}
\equiv \frac{\rm RM}{\rm DM},
\end{equation}
where $B_{\parallel} \equiv B \cos\theta$ and $\theta$ is the angle
between ${\bmit B}$ and the line of sight.
Several authors \citep[e.g.,][]{han98,id99,fr01,han06,be07} have used
the RMs and DMs of pulsars to reproduce the large-scale magnetic field
of our Galaxy.

Recently, the distributions of RMs along many contiguous lines of
sight have been obtained in multi-frequency polarimetric observations
of the diffuse Galactic synchrotron background \citep{ha03,ha04,sc09}
and the Perseus cluster \citep{bb05}.
While the peak in the frequency distribution of the RMs reflects the
regular component of the magnetic field, the spread should reflect
the turbulent component.
This means that if the distribution of RMs is observed, the spread
provides another way to quantify the magnetic field in turbulent
ionized media such as the WIM.

Motivated by the importance of RM in the study of the magnetic field
in the WIM, in this paper we study RM in turbulent media with the rms
Mach number of unity.
Simulations are outlined in Section 2.
Results are presented in Sections 3 and 4.

\section{Simulations}

We performed three-dimensional simulations of isothermal, compressible,
magnetohydrodynamic (MHD) turbulence, using a code based on the total
variation diminishing scheme \citep{kim99}.
The fact that the temperature of the WIM is in a relatively narrow range of
6000 - 10000 K \citep{ha99} would justify the assumption of isothermality.
There are two parameters in our simulations: the root-mean-square (rms)
sonic Mach number, $M_s$, and the initial plasma beta value, $\beta_0$.
Hereafter, the subscript ``0'' is used to denote the initial values.
The first parameter indicates the level of turbulence.
In this paper, we focus on the turbulence with the rms Mach number of
unity, and so set $M_s = 1$.
The second parameter tells the strength of the initially uniform magnetic
field, or the regular field, $B_0$.
In order to explore the effect of regular field, we varied the value of
$\beta_0$ from 0.1, 1, 10, to 100.
If we take 8000 K and 0.03 cm$^{-3}$ as the representative values of
the temperature and electron density, assuming that hydrogen is
completely ionized, helium is neutral, and the number ratio of
hydrogen to helium is 0.1, we have $B_0 = 1.3 (1/\beta_0)^{1/2}
(T/8000~{\rm K})^{1/2} (n_e/0.03~{\rm cm}^{-3})^{1/2}~\mu$G.
So the initial magnetic field strength in our simulations corresponds to
4.1, 1.3, 0.41, 0.13 $\mu$G for $\beta_0$ = 0.1, 1, 10, and 100,
respectively.

Simulations were started with $B_0$ along the $x$-direction in a uniform
medium of density, $\rho_0$.
The grid of $512^3$ zones was used for the periodic computational box of
size $L$.
Turbulence was driven with the recipe in \citet{st99} and \citet{ml99};
velocity perturbations were drawn from a Gaussian random field
determined by the top-hat power distribution in a narrow wavenumber
range of $(2\pi/L)\leq k\leq 2 (2\pi/L)$, and added at every
$\Delta t = 0.001 L/c_s$.
The amplitude of the perturbations was tuned in such a way that
$M_s \equiv v_{\rm rms} /c_s$ became nearly unity at saturation.
Here, $c_s$ is the isothermal sound speed and $v_{\rm rms}$ is the
rms velocity of the resultant turbulent flow.
In simulations $v_{\rm rms}$ initially increased and became saturated
at $t \simeq (1/2) L/c_s$, and we ran the simulations up to $t = 2 L/c_s$,
two sound crossing times.
At saturation, the average strength of magnetic field reached 4.2,
1.45, 0.73, and 0.51 $\mu$G for $\beta_0$ = 0.1, 1, 10, and 100,
respectively.
The amplification factor was larger for the simulations with larger
$\beta_0$ or weaker $B_0$.

\section{Correlation between $B$ and $\rho$}

We first check whether $\langle B_{\parallel}\rangle$ in Equation (1)
correctly reproduces the unbiased strength of the magnetic field along
the line of sight.
That is true, only if there is no correlation between $B$ and $n_e$
(or the gas density $\rho$).
If the correlation is positive, $\langle B_{\parallel}\rangle$ in
Equation (1) overestimates the magnetic field strength, while if
negative, it underestimates the strength.

It is known from observations \citep{crut99,pn99} and numerical
simulations \citep[e.g.,][]{os01,pas03,bk05,ml05,bu09} that in the
highly compressible, supersonic, molecular cloud environment,
the magnetic pressure (or $B$) and the gas pressure (or $\rho$ in
isothermal gas) are positively correlated.
\citet{bu09}, on the other hand, reported a weak negative correlation
for $M_s = 0.7$ and $\beta_0 = 2$ (their model 1).
Figure 1 shows the correlation between $B$ and $\rho$ in our simulations.
The nearly circular shape of contour lines indicates weak correlation
for $M_s = 1$.
To quantify it, we calculated the correlation coefficient
\begin{equation}
r(B,\rho)=
\frac{{\Sigma}_{i,j,k} (B_{i,j,k}-\bar B)(\rho_{i,j,k}-\bar \rho)}
{\left[\Sigma_{i,j,k}(B_{i,j,k}-\bar B)^2\right]^{1/2}
\left[\Sigma_{i,j,k}(\rho_{i,j,k}-\bar \rho)^2\right]^{1/2}},
\end{equation}
where $\bar B$ and $\bar \rho$ are the average values of $B$ and $\rho$.
The values of $r$ at the end of simulations were 0.01, -0.12, -0.06,
and 0.16 for $\beta_0$=0.1, 1, 10, and 100, respectively.
Relatively small values of $r$ confirm that the correlation is weak for
$M_s = 1$, regardless of $\beta_0$.

Weak correlation means that the RM field in Equation (1) should correctly
represent the true magnetic field.
To test it, we compare the strengths of two fields in Figure 2;
the frequency distributions of the true mean strength of the magnetic
field along the line of sight (presented with solid lines) and the
magnetic field strength calculated with Equation (1) (presented with
dotted lines) coincide quite well.
So we argue that the systematic bias, due to a correlation between $B$
and $n_e$, in the estimation of magnetic field strength with RM would
be insignificant in the turbulent media with $M_s = 1$.

\citet{be03} pointed that the discrepancy in the strengths of the
regular Galactic magnetic field estimated with RM and synchrotron
emissivity could be reconciled, if the correlation between $B$ and
$n_e$ is negative in the WIM and so the RM field was underestimated.
\citet{be03} postulated the pressure equilibrium, which results in the
negative correlation between $B$ and $\rho$.
If the pressure equilibrium is maintained, the combined pressure of
gas and magnetic field, $P_{\rm tot} = P_{\rm gas} + P_{\rm mag}$,
should exhibit a narrow distribution.
Figure 3 shows the frequency distribution of $P_{\rm tot}$.
Our simulations show broad distributions rather than peaked
distributions.
That is, our results do not support the prediction of \citet{be03}.
However, we caution that we considered only the turbulent media with
the rms Mach number $M_s = 1$.
On the other hand, the WIM may have $M_s$ which is not exactly unity.
So we need to further investigate turbulent media with $M_s \ne 1$,
before we exclude the postulation of \citet{be03}.
We leave it as a future work.

\section{Frequency Distribution of RMs}

We then look at the frequency distribution of RMs.
Figure 4 shows the probability distribution of RM/$\overline{\rm
RM}$ for different $\beta_0$'s as well as for different $\theta$'s.
Here, $\overline{\rm RM}$ is the average value of RMs. 
(The distribution was also calculated for $\theta = 83^\circ$, but
not shown in the figure for clarity.
The distribution for $\theta = 83^\circ$ is used in Figure 5.)
The fit to the Gaussian is also shown.
A noticeable point in the figure is that the distribution of
RM/$\overline{\rm RM}$ is very well fitted to the Gaussian.
The goodness-of-fit is between 0.89 and 0.99
\footnote{We used the coefficient of determination, $R^2$, as an
indicator of goodness-of-fit, where $R^2 = 1$ if the fit is perfect
and $R^2 < 1$ if the fit is less perfect.}.
This result is consistent with the observations of \citet{ha03,ha04}.
They took the multi-frequency polarimetric images of
diffuse radio synchrotron backgrounds in the constellation Auriga
and Horologium, and obtained RM maps.
The distribution of the observed RMs is also fitted to the Gaussian.

Another noticeable point in Figure 4 is that the distribution is
more widely spread for larger $\beta_0$ and larger $\theta$.
It hints at a possible correlation between the spread in the
distribution of RM/$\overline{\rm RM}$ with the strength of the
magnetic field along the line of sight.
Figure 5 shows the relation between the full width at half maximum,
$W_{\rm FWHM}$, of the distribution of RM/$\overline{\rm RM}$ shown in
Figure 4 and the strength of the regular field along the line of sight,
$B_{0\parallel} \equiv \ B_0 \cos\theta$.
The best fit for the relation
\footnote{We note that in the fit, there is a trend that the blue symbols
for $\beta_0 = 0.1$ are mostly above the fitting line, while the black
symbols for $\beta_0 = 100$ are mostly below the line.
This tells that $\beta_0$ may enter the relation as a secondary,
but less prominent, parameter.}
is
\begin{equation}
B_{0\parallel} = (2.45\pm0.3) \times W_{\rm FWHM}^{-1.41\pm0.1}~~\mu {\rm G}.
\end{equation}
Note that the relation is for $T = 8000$ K and $n_e = 0.03$ cm$^{-3}$
as the representative values of the temperature and electron density,
and  scales as $(T/8000~{\rm K})^{1/2} (n_e/0.03~{\rm cm}^{-3})^{1/2}$
for other values of $T$ and $n_e$.
The empirical relation in (3) may provide a handy way to quantify the
strength of the regular field along the line of sight in regions where
the map of RMs has been obtained along many contiguous lines of sight
with the multi-frequency polarimetric observations (see Introduction).
The accompanying map of DMs is not necessary for it.

The estimation of magnetic field strength with the relation in (3),
however, should be done with caution because of its limitations:
First, the relation is valid only for $M_s=1$.
Second, no stratification of gas and magnetic field was assumed, which
is apparent in the polarization maps of a larger area of our Galaxy. 
Third, no source emitting polarized light in the medium between
background polarized continua and us was considered.

From a physical point of view, the broadening of the width of RM
distribution is due to fluctuating magnetized gas.
We can easily expect that, for a given regular field, the width would be
larger if the rms Mach number, $M_s$, is larger.
So $W_{\rm FWHM}$ should be a function of not only $B_{0\parallel}$
but also $M_s$.
In order to quantify the relation of $W_{\rm FWHM}$ versus $B_{0\parallel}$
and $M_s$, we need simulations that cover the parameter space of $\beta_0$
and $M_s$.
Low-resolution simulations with $M_s=0.5$ and 2.0 show that the
probability distribution of RM/$\overline{\rm RM}$ is still fitted
to the Gaussian, and $W_{\rm FWHM}$ can differ by a factor of two to
three if $M_s$ differs by a factor of two.
We leave the report of the $M_s \ne 1$ results, including the correlation
between $B$ and $\rho$, from high-resolution simulations as a future work.

Nevertheless, we can try to apply the relation in (3) to RM observations
in the WIM, assuming $M_s = 1$ there.
For instance, from the figures of the frequency distribution of RMs in
Auriga and Horologium in \citet{ha03,ha04}, we read $W_{\rm FWHM} \simeq 2$
and 7, and get $B_{0\parallel} \sim 0.6$ and 0.1 $\mu$G for Auriga
and Horologium, respectively (if $n_e = 0.016$ cm$^{-3}$ is used as in
\citet{ha04}).
These are close to, but somewhat larger than, the values estimated in
\citet{ha04}, $\sim 0.42$ and 0.085 $\mu$G, respectively.
However, by considering the limitations of the relation in (3) and the
uncertainty in the values of $W_{\rm FWHM}$ which we read from
figures, we regard the agreement to be fair.

Currently the number of the synchrotron backgrounds, which can be used
for the observation of RM maps, is still limited due to the relatively
low sensitivity of present-day radio telescopes.
However, the new-generation radio telescopes, such as LOFAR (Low Frequency
Array) and SKA (Square Kilometer Array) with much higher sensitivity, will
certainly provide RM maps in much larger portions of the sky.
Then, the relation like the one in (3) will become a useful diagnosis.

\acknowledgments

We thank R. Beck and anonymous referees for clarifying comments.
The works of JK, DR and JC were supported by the Korea Foundation
for International Cooperation of Science and Technology through
K20702020016-07E0200-01610.
The work of JK was also supported by National Research Foundation of
Korea through 2009-0062863 (ARCSEC).
The work of DR was also supported by Korea Science and Engineering
Foundation (KOSEF) through R01-2007-000-20196-0.
This work utilized a high performance cluster built with fundings
from the Korea Astronomy and Space Science Institute and KOSEF
through the Astrophysical Research Center for the Structure and
Evolution of Cosmos (ARCSEC).

\clearpage

\begin{figure}
\vskip -5.cm
\hskip -5.cm
\includegraphics[scale=1.4]{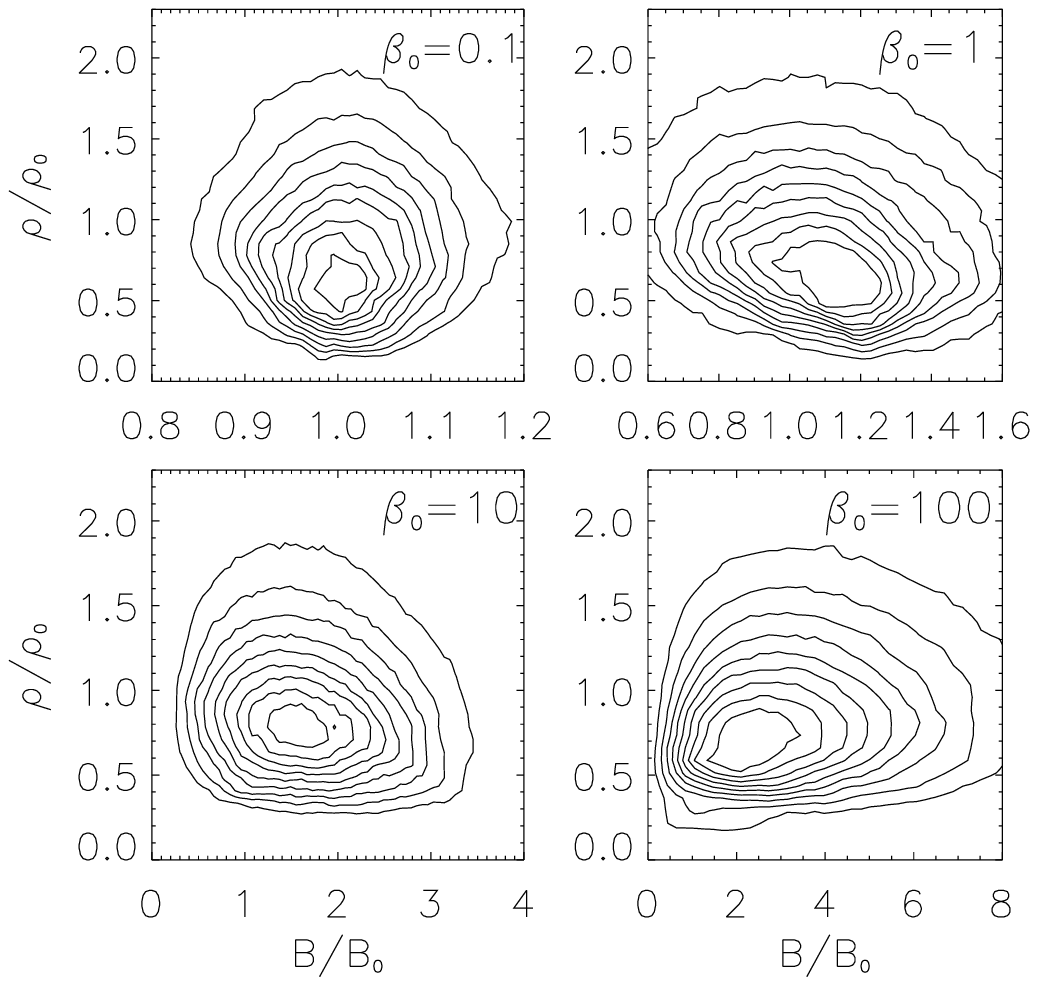}
\vskip -6.cm
\caption{Contour plots showing the correlation between the magnetic
field strength and the gas density at the end of fours simulations.
The contour levels are from 10\% to 90\% of the peak value with
the interval of 10\%.}
\end{figure}

\begin{figure}
\vskip -0.9cm
\hskip -3.4cm
\includegraphics[scale=1.1]{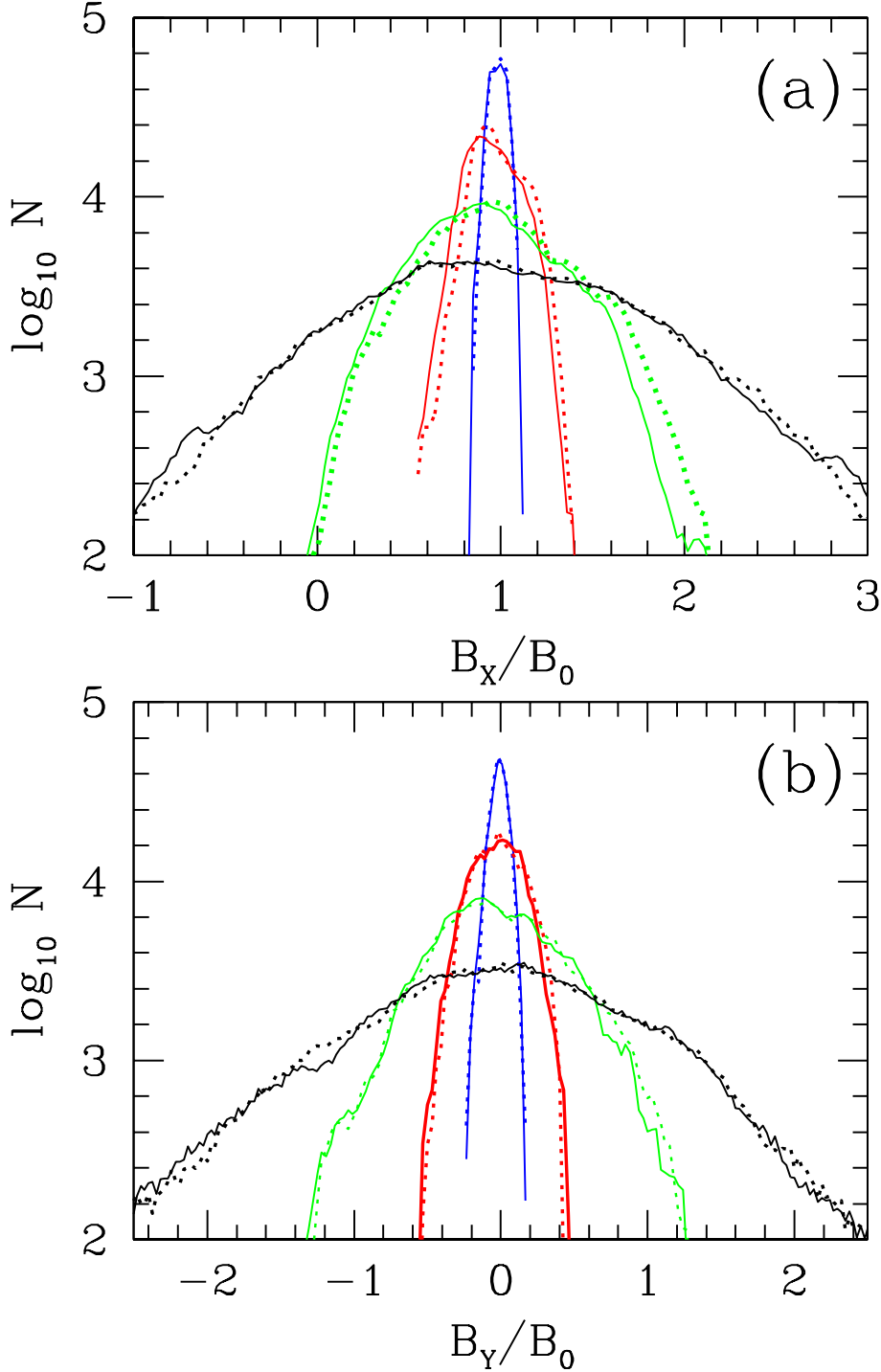}
\vskip -2.8cm
\caption{Frequency distribution of (a) $\langle B_x \rangle$ and (b)
$\langle B_y \rangle$, normalized with the initial magnetic field strength,
$B_0$.
The solid lines present the true mean strength of the magnetic field along
the line of sight, and the dotted lines present the magnetic field strength
calculated with RM/DM, along $512^2$ lines aligned the (a) $x$ and (b)
$y$-axes.
The blue, red, green, and black lines are for the simulations with $\beta_0$
= 0.1, 1, 10, and 100, respectively.}
\end{figure}

\begin{figure}
\vskip -2.cm
\hskip -2.8cm
\includegraphics[scale=1.]{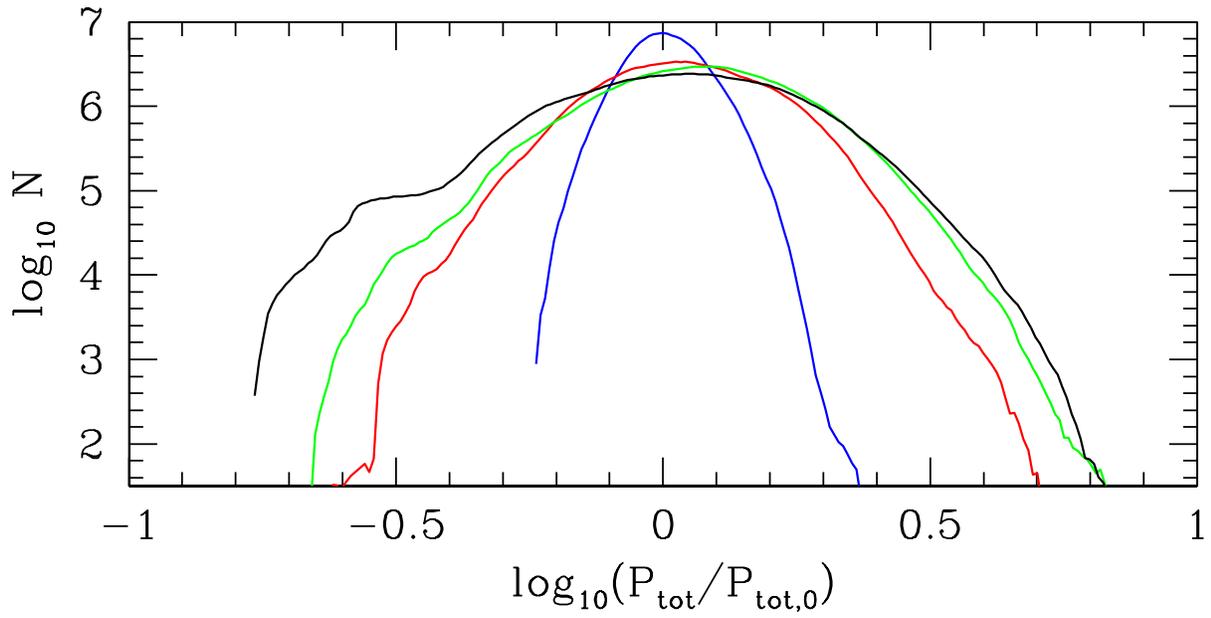}
\vskip -7.5cm
\caption{Frequency distribution of the total (gas and magnetic) pressure.
The blue, red, green, and black lines are for the simulations with $\beta_0$
= 0.1, 1, 10, and 100, respectively.}
\end{figure}

\begin{figure}
\vskip -0.6cm
\hskip -2.cm
\includegraphics[scale=1.]{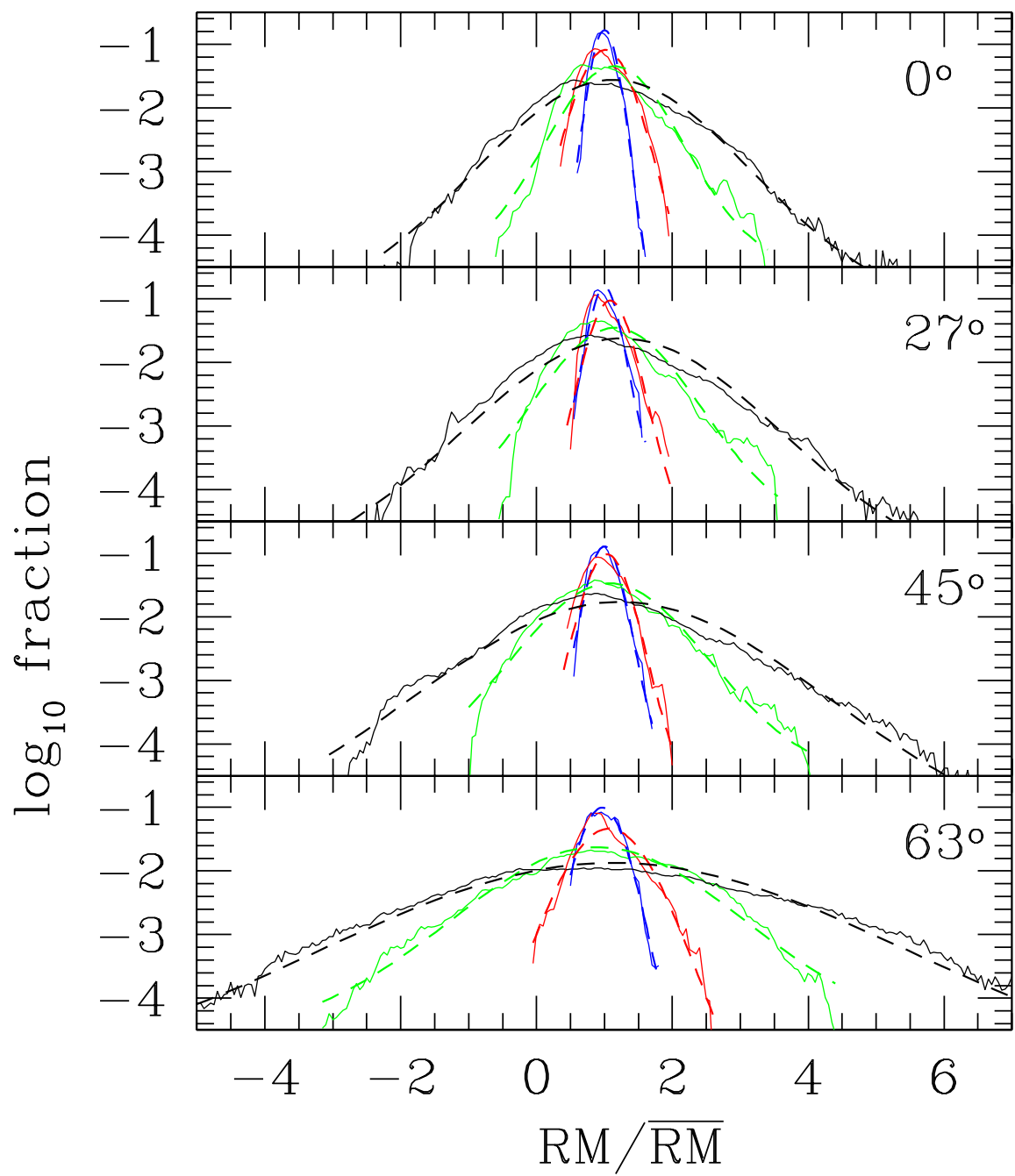}
\vskip -1.6cm
\caption{Probability distribution of RMs normalized with the average
value.
The solid lines are the distribution, and the corresponding dashed lines
are the Gaussian fit.
The blue, red, green, and black lines are for $\beta_0$ = 0.1, 1, 10,
and 100, respectively.
Different panels are for different values of $\theta$ (the angle
between the direction of the regular field and the line of sight).}
\end{figure}

\begin{figure}
\vskip -3.cm
\hskip -0.5cm
\includegraphics[scale=0.85]{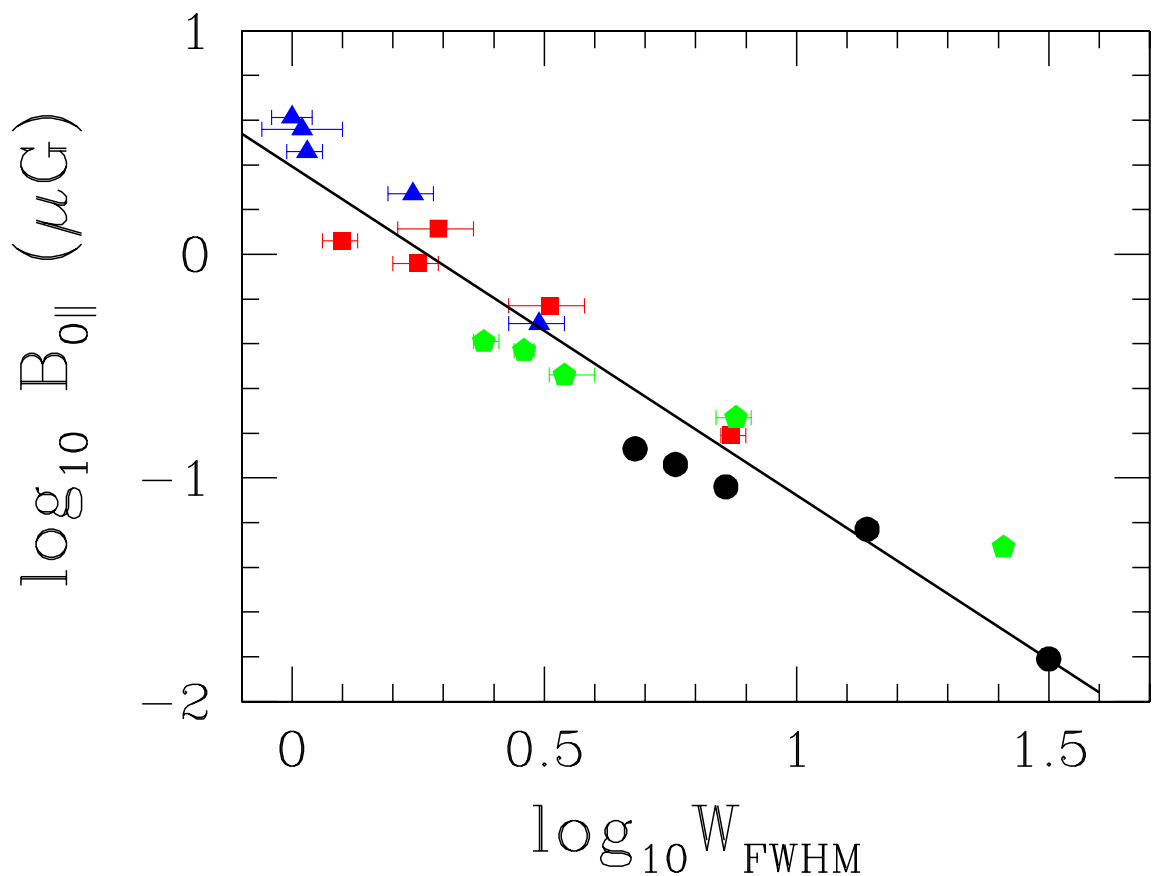}
\vskip -3.cm
\caption{Relation between the full width at half maximum of the distribution
of RM/$\overline{\rm RM}$ ($W_{\rm FWHM}$) and the strength of the regular
field along the line of sight ($B_{0\parallel}$).
Blue, red, green, and black symbols are from the simulations with
$\beta_0$ = 0.1, 1, 10, and 100, respectively (for five different $\theta$'s
including $\theta = 83^\circ$).
The error bar shows the error in the Gaussian fit in Figure 4.
The solid line shows the best fit of the relation between $W_{\rm FWHM}$
and $B_{0\parallel}$.}
\end{figure}

\end{document}